\documentclass[12pt,preprint]{aastex}
\usepackage{graphicx}

\newcommand{\hst}{{\it HST\/}}

\slugcomment{Accepted by the Astronomical Journal, on March 21, 2012} 

\shorttitle{outer field in Omega Cen}
\shortauthors{King et al.}
\usepackage{ulem}

\begin{document}

\def\subr #1{_{{\rm #1}}}


\title{{\it Hubble Space Telescope} Observations of an Outer Field in
 Omega Centauri:\ A Definitive Helium Abundance\footnote{Based on
   observations with the NASA/ESA {\it Hubble Space Telescope}, obtained
   at the Space Telescope Science Institute, which is operated by AURA,
   Inc., under NASA contract NAS 5-26555.}}
\author{
I.\ R.\ King\altaffilmark{1},
L.\ R.\ Bedin\altaffilmark{2,6},
S.\ Cassisi\altaffilmark{3},
A.\ P.\ Milone\altaffilmark{4},
A.\ Bellini\altaffilmark{2}
G.\ Piotto\altaffilmark{5},
J.\ Anderson\altaffilmark{2},
A.\ Pietrinferni\altaffilmark{3}, and
D.\ Cordier\altaffilmark{7}
}

\altaffiltext{1}{Department of Astronomy, University of Washington,
Box 351580, Seattle, WA 98195-1580; king@astro.washington.edu}

\altaffiltext{2}{Space Telescope Science Institute, 3700 San Martin
Drive, Baltimore, MD 21218; [bellini;jayander]@stsci.edu}

\altaffiltext{3}{Osservatorio Astronomico di Teramo, Via Mentore Maggini
  s.n.c., I-64100 Teramo, Italy; [cassisi;adriano]@oa-teramo.inaf.it}

\altaffiltext{4}{Instituto de Astrof\`\i sica de Canarias, E-38200 La
Laguna, Tenerife, Canary Islands, Spain; Department of Astrophysics,
University of La Laguna, E-38200 La Laguna, Tenerife, Canary Islands,
Spain; milone@iac.es}

\altaffiltext{5}{Dipartimento di Astronomia, Universit\`a di Padova,
Vicolo dell'Osservatorio 2, I-35122 Padova, Italy;
giampaolo.piotto@unipd.it}

\altaffiltext{6}{INAF-Osservatorio Astronomico di Padova, Vicolo
  dell'Osservatorio 5, I-35122 Padova, Italy; luigi.bedin@oapd.inaf.it}

\altaffiltext{7}{Institut UTINAM, Observatoire de Besançon - UMR CNRS
  6213, 41 bis avenue de l’Observatoire, F-25000 Besançon, France;
  daniel.cordier@obs-besancon.fr}

\begin{abstract}

We revisit the problem of the split main sequence (MS) of the globular
cluster $\omega$~Centauri, and report the results of two-epoch {\it
Hubble Space Telescope} observations of an outer field, for which proper
motions give us a pure sample of cluster members, and an improved
separation of the two branches of the main sequence.  Using a new set of
stellar models covering a grid of values of helium and metallicity, we
find that the best possible estimate of the helium abundance of the
bluer branch of the MS is $Y=0.39\pm 0.02$.

For the cluster center we apply new techniques to old observations:\ we
use indices of photometric quality to select a high-quality sample of
stars, which we also correct for differential reddening.  We then
superpose the color-magnitude diagram of the outer field on that of the
cluster center, and suggest a connection of the bluer branch of the MS
with one of the more prominent among the many sequences in the subgiant
region.  We also report a group of undoubted cluster members that are
well to the red of the lower MS.

\end{abstract}


%
\section{Introduction}
%
%

The stellar system $\omega$ Centauri (NGC 5139) is one of the most
puzzling objects in our Galaxy. In an intriguing way, the large amount
of attention devoted to this globular cluster has served to increase
rather than decrease the number of mysteries that surround it.  For
several decades it had been recognized that the red giant branch (RGB)
of $\omega$ Cen shows a spread in metallicity (Dickens \& Woolley 1967,
Cannon \& Stobie 1973).  The breadth in color of the RGB was interpreted
as an indication of an intrinsic spread in chemical abundance, as
subsequently confirmed by spectroscopic data (Freeman \& Rodgers 1975);
and Norris, Freeman, \& Mighell (1996) suggested that two epochs of
star formation have occurred.  Today we know that there are many more
populations than that (Lee et al.\ 1999, Pancino et al.\ 2000, Sollima
et al.\ 2005, Villanova et al.\ 2007, Bellini et al.\ 2010, and
references therein).  It has been suggested further that at each
metallicity there is a range in age (Sollima et al.\ 2005, Villanova et
al.\ 2007).

More than a decade ago our group found that the main sequence (MS) of
the cluster is double (Anderson 1997, 2002; Bedin et al.\ 2004).
Because of the lack of information on the heavy elements in the MS
components, the original version of Bedin et al.\ (2004) had proposed
several explanations for the split, none of them conclusive.  The
referee, John Norris, strongly suggested, however, ``TRY HELIUM'', in
spite of the improbably high He abundance that would be required, and
published his suggestion (Norris 2004).

The turning point was the work by Piotto et al.\ (2005), who actually
measured the heavy elements along the two branches of the MS.  They
found that contrary to expectation, the bluer branch (bMS) has a higher
rather than a lower metallicity than the redder branch (rMS), and that
the only way to explain the photometric and spectroscopic results was
that the helium abundance of the bMS is $Y \sim$ 0.4 --- far beyond what
its metallicity would imply.  Very recently Dupree, Strader, \& Smith
(2011) have found direct evidence for an enhancement in He, from the
analysis of the $\lambda$10830 transition of He I in the RGB stars of
$\omega$ Cen.  The correlation of He-line detection with [Fe/H], Al, and
Na supports the assumption that He is enhanced in stars of the bMS.  (It
is still totally unclear, however, where a large enrichment of He could
have come from, and how it could fit with current stellar-evolution
models.)

In other clusters too, studies of the RGB (Bragaglia et al.\ 2010a)
and of the HB (Gratton et al.\ 2010) have suggested a spread in He
abundances.  In both of these studies NGC 2808 has stood out
especially as showing a clear spread in helium --- not surprisingly,
since that cluster has a main sequence that is split into three
branches (Piotto et al.\ 2007).  And in that cluster Pasquini et
al.\ (2011) have used the He $\lambda$10830 line to estimate a helium
value $Y\geq0.39$, confirming suggestions by D'Antona \& Caloi (2008)
and by Bragaglia et al.\ (2010a); furthermore, Bragaglia et
al.\ (2010b) have obtained spectra of MS stars on the reddest and on
the bluest MS branches of NGC 2808, and find that abundances of
individual heavy elements are very much in agreement with the scenario
of normal He for the rMS and enhanced He for the bMS.

In the present paper we describe an outer field of $\omega$ Cen that has
relatively few stars, but minimal crowding.  This field has already been
used in our paper on the splitting of the MS (Bedin et al.\ 2004), and
also in a study of the radial behavior of the numbers of stars in the
bMS and the rMS, by Bellini et al.\ (2009); the present paper is the
fuller discussion that the latter authors promised for this field.  In
addition to our presentation of the MS split in this outer field, we
sharpen our view of the central field, we fit theoretical isochrones to
the two branches of the main sequence, and, importantly, we estimate the
difference in their helium abundances, along with the quantitative
uncertainty of that difference.

%
\section{The Outer Field} 
\label{outer}
%

\subsection{Observations, Measurements, and Reductions} 
\label{obs}

In the outer field (13$^{\rm h}$ 25$^{\rm m}$ 35\fs5,
$-$47\arcdeg\ 40\arcmin\ 6\farcs7, same as the 17\arcmin\ field in Bedin
et al.\ 2004) we combined a new data set with an earlier one.  We had
already imaged this field, 17$^{\prime}$ from the center of $\omega$~Cen
(core radius 1\farcm 4, half-mass radius 4\farcm 2, tidal limit
57$^{\prime}$), using the Wide Field Channel (WFC) of the {\it Hubble
  Space Telescope's} (\hst 's) Advanced Camera for Surveys (ACS).  Those
images (GO-9444, PI King), taken 3 July 2002, consisted of 2$\times$1300
s + 2$\times$1375 s with F606W, and 2$\times$1340 s + 2$\times$1375 s
with F814W.  Our follow-up program (GO-10101, PI King) was to have had
second-epoch images in F814W only, but in view of the extreme interest
ignited by the results presented in Bedin et al.\ (2004) and Piotto et
al.\ (2005), we were able to get Director's Discretion time that allowed
us to repeat the F606W images as well.  After a delay caused by a failed
guide star, the second-epoch images were taken 24 Dec 2005, with
exposures 2$\times$1285 s + 2$\times$1331 s in F606W and 4$\times$1331 s
in F814W.  Because of the delay, however, the orientation differed by
180 degrees.  As we shall see, this change improved the photometry, but
at the cost of somewhat complicating the astrometry.

The photometry was carried out using the procedures and software tools
developed for the Globular Cluster Treasury program (Sarajedini et al.\
2007), as described in detail by Anderson et al.\ (2008).  We summarize
here briefly: To each star image in each exposure we fit a point spread
function (PSF) interpolated expressly for that star, using a 9 $\times$
5 array of PSFs in each of the two chips of the ACS/WFC.  These arrays
model the spatial variation of the PSF, but for each individual exposure
we add a ``perturbation PSF'' that fine-tunes the fitting to allow for
small differences in focus, temperature, etc.

The fitting of each star uses its central 5 $\times$ 5 pixels, and
yields a flux and a position.  In addition we created a model of the
extended outer parts of the PSF that allowed us to eliminate the
artifacts that arise from outer features in the PSFs of bright stars,
while excluding very few legitimate stars. (The PSFs are described
in great detail in Anderson et al.\ 2008.)

We transformed our zero points to those of the WFC/ACS Vega-mag system
following the procedure given in Bedin et al.\ (2005), and using the
encircled energy and zero points given by Sirianni et al.\ (2005).
Because of the high background ($\gtrsim$100 $e^{-}$ pixel$^{-1}$),
combined with the fact that our exposures were taken at a time when
inefficiencies in charge transfer were less serious than they are now,
this field did not need any corrections for inadequate charge-transfer
efficiency.

\subsection{The Saturated Stars}
\label{SAT}

Since the primary aim of the programs for which the images were taken
had been the faint stars, neither of our epochs included short
exposures.  For the bright stars we had to derive the best photometry
that we could from their saturated images.  We used the method developed
by Gilliland (2004), which works by recovering electrons that have bled
into neighboring pixels.  Our application of this method is described in
Sect.\ 8.1 of Anderson et al.\ (2008).  The method depends, however, on
having a detector GAIN greater than 1.  Unfortunately our first-epoch
images were taken with GAIN=1, but in the second epoch we had GAIN=2 and
were able to do effective photometry on the saturated images.  Thus for
saturated stars we had only the second epoch available.

\begin{figure}[!ht]
\epsscale{1.00}
\plotone{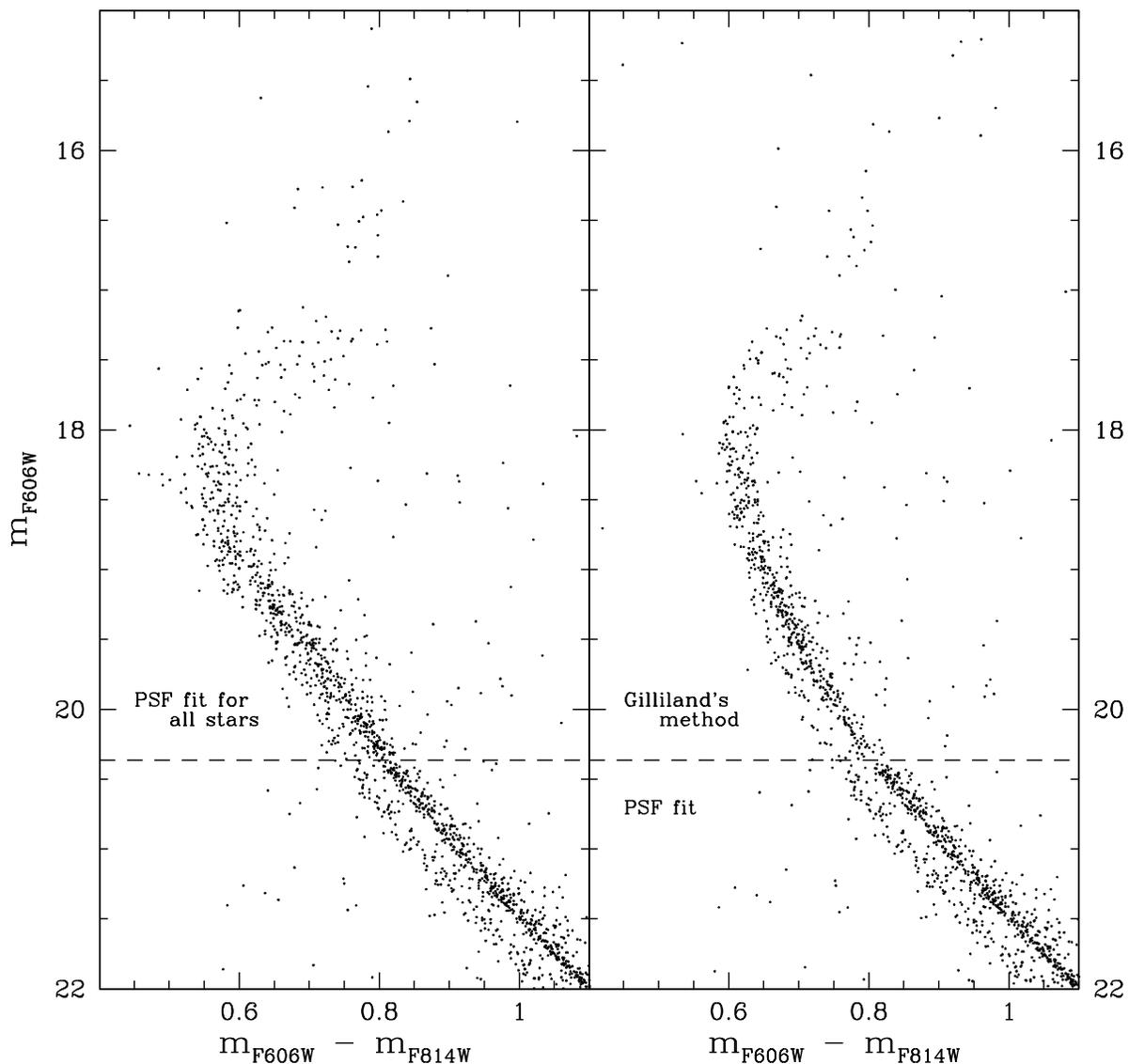}
\caption{(Left) The CMD as obtained from PSF fitting of unsaturated
pixels. The horizontal dashed line marks the beginning of saturation in
filter F606W.  (Right) The same CMD, with photometry for saturated stars
obtained as described in Sect.~\ref{SAT}.}
\label{sat}
\end{figure}

Because Gilliland's method uses so different a procedure, we had to
adjust the zero points of the magnitudes that it produced.  The
necessary shifts ($<0.02$ mag in each band) were easily determined from
the stars in the 0.5-mag interval just below the saturation limit, where
both methods are valid.  Figure \ref{sat} shows the impressive
improvement that the use of Gilliland's method gave us for a stretch of
about 3 magnitudes at the bright end.  It also shows, however, that for
stars brighter than $m_{\rm F606W}\sim 17.1$ even that method is unable
to cope with saturation.  (From the onset of saturation at
magnitude 20.4 up to a magnitude about 17.1, however, there is a
considerable improvement.)

\subsection{Proper Motions}
\label{ast}

For the astrometry we did not need the elaborate procedures that we had
used for the photometry; for our data set it was more appropriate simply
to use the program described by Anderson \& King (2006), to get position
coordinates for each star in each {\tt \_flt} image.  (These are
the images that have been bias-subtracted and flat-fielded via the
standard ACS pipeline, but have not been resampled; as such they are
suitable for derivation of positions and fluxes by high-precision
PSF-type analysis.)

Our next step was to apply the distortion corrections of Anderson
(2002, 2006).  We then needed to transform these positions into a common
reference frame at each epoch, for which we arbitrarily chose one image
at that epoch.  For the transformations we selected among the brighter
stars a set of stars that were in the main-sequence region of the
color-magnitude diagram (CMD), so as to minimize any disturbing
influence that inadvertent inclusion of field stars might have on our
transformations.

Once we had all the positions in a single reference frame at each epoch,
we measured the displacement of each star from the first epoch to the
second, relative to a set of at least 10 of its immediate neighbors
(dropping the few stars that did not have 10 near neighbors), so that
each cluster star would have a near-zero displacement between the two
epochs, while field stars would show noticeable motions.

Since the astrometry depends very little on the filter band, we were
able to treat each filter the same, and combine the results.  With a
precision of better than 0.05 pixel for the coordinates of individual
star images, the precision of a displacement from the mean of 8
exposures at each epoch should be 1/2 of that, so that over the 3.5-year
baseline the proper motions should be good to 0.007 pixel per year,
which for a 50-milliarcsec pixel is 0.35 mas yr$^{-1}$.  Since the
distance to $\omega$ Cen is about 5 kpc, this is equivalent to an
uncertainty of $\sim 4$ km$^{-1}$ in the transverse motion of each star
(appreciably less than the 10 km sec$^{-1}$ internal velocity dispersion
of the cluster at this distance from the center [Merritt, Meylan, \&
Mayor 1997]).  In practice, moreover, the field-star motions turn out to
differ from the cluster motions by much more than the internal
dispersion of the latter.

\begin{figure}[!hb]
\epsscale{1.00} \plotone{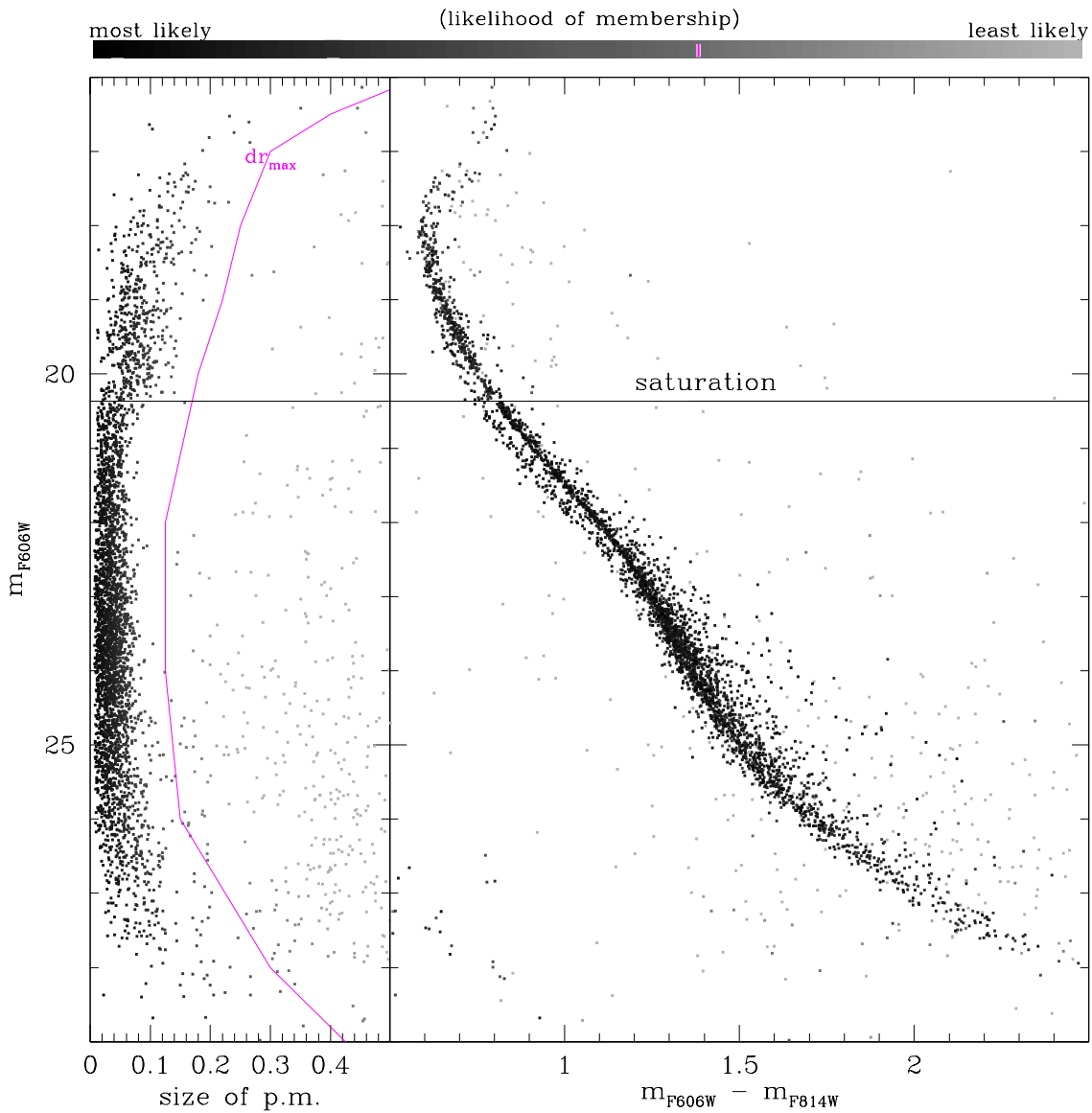}
\caption{At left, the sizes of the proper motions (in pixel units),
  with the symbol for each star gray-scale-coded so as to show in an
  intuitive way the likelihood that it is a cluster member (as
  indicated in the palette at the top).  The magenta lines in this
  panel and in the palette show our proper-motion cut-off for
  membership.  In the right hand panel is the CMD of the same stars,
  with the same gray-scale coding, so that one can see the membership
  likelihood of each star in the CMD.}
\label{memb}
\end{figure}

\subsection{The Cleaned Color-Magnitude Diagram of the Outer Field}
\label{dCMD}

Because this field is so far from the center of $\omega$ Cen, it suffers
relatively greater contamination by field stars than inner fields do; we
therefore made careful use of proper motions to remove field stars from
our color-magnitude diagram.  Our first step was to plot the $x$ and $y$
components of proper motion for each unit interval of magnitude.  From
these plots it was clear that most of the motions --- those of cluster
members --- were concentrated around a common centroid, while the
motions of field stars scattered much more widely about another center.
For bright stars there was little or no overlap between the two
distributions, but with increasing measurement error at faint magnitudes
the separation became less clear.

\begin{figure}[!ht]
\epsscale{1.00} \plotone{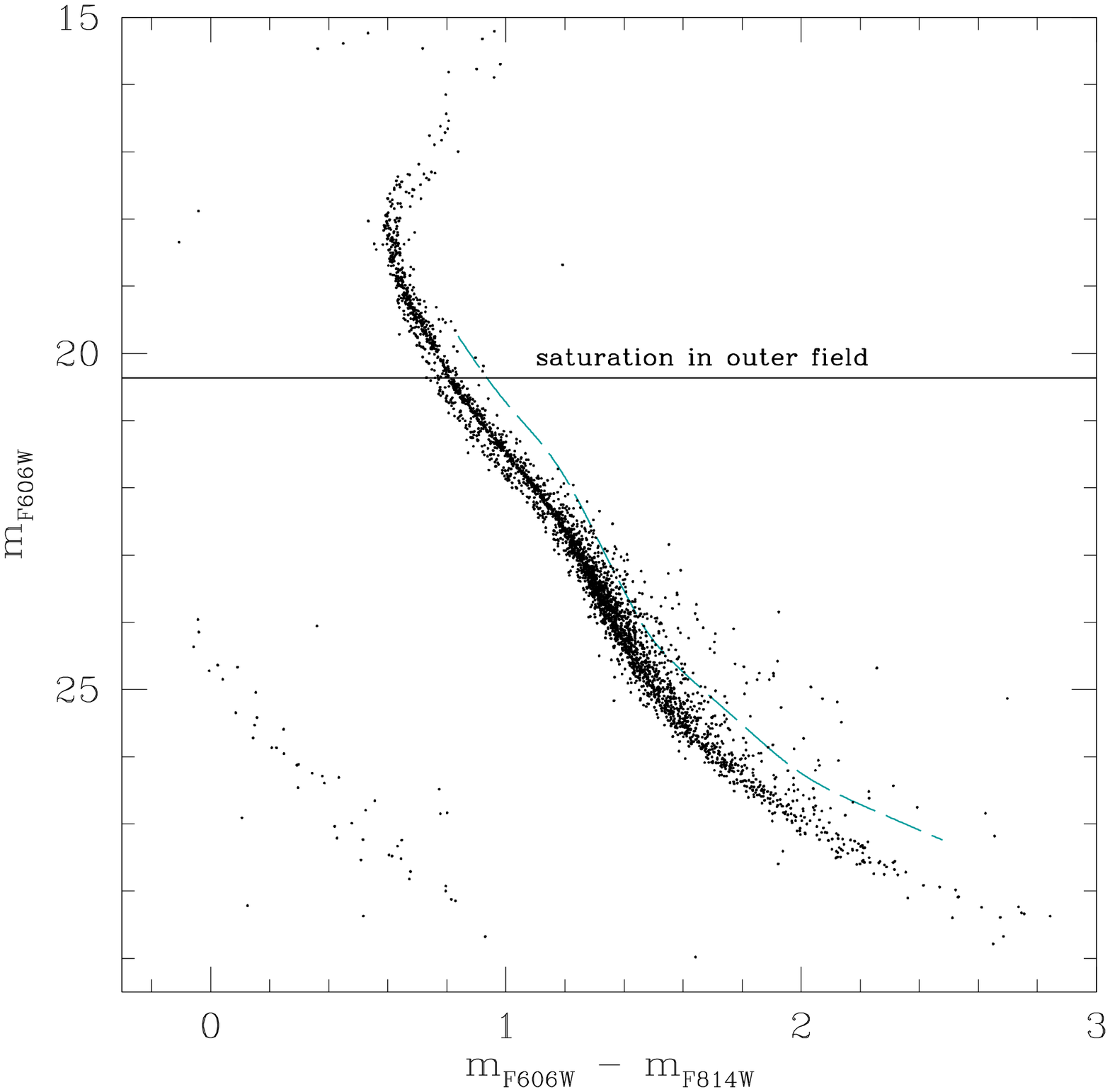}
\caption{The color-magnitude diagram of cluster members in the outer
  field (stars whose proper motions lie to the left of the magenta line
  in the preceding figure).  The dashed blue line marks the locus
  of equal-mass rMS binaries, 0.75 mag above the rMS.}
\label{outfig}
\end{figure}

In the left-hand panel of Figure \ref{memb} we show the sizes of the
motions as a function of magnitude.  We drew an arbitrary line, shown in
magenta, to separate members from non-members.  Rather than just
throwing away the stars that we call non-members, however, we have
colored each star from black to gray according to the palette shown at
the top of the figure.  Stars whose motions are very close to that of
the cluster centroid are black; with increasing difference from the mean
cluster motion the symbols become a paler and paler gray.

With the likelihood that each star is a cluster member coded in this
intuitive way, in the right-hand panel of the figure we plot the stars
in the color-magnitude diagram.  The symbol that represents each star
now has the same degree of grayness that the star was assigned in the
left-hand panel, so that we can see where in the CMD the cluster members
and the non-members lie, and conversely, from the gray level of the
symbol, which stars should be rejected as non-members.  Finally, in
Figure \ref{outfig} we show the CMD of the stars whose proper motions
lie to the left of the magenta line in the preceding figure, which we
therefore consider to be cluster members.  The dashed blue line, 0.75
mag above the rMS, marks the upper limit of its binaries (those with
equal mass).

As we have already noted, the flaring out of the MS at magnitudes
brighter than 17 is due to the inability of our methods to cope with the
most extreme levels of saturation.  This anomaly aside, our CMD shows a
number of interesting features: (1) The main sequence shows the best
separation of its two branches that has yet been seen.  (It should be
noted that Fig.\ \ref{outfig} is similar to Fig.\ 1(d) of Bedin et
al.\ 2004, but now the interfering field stars have been removed.  For
the relative number of rMS and bMS stars one should see Bellini et
al.\ 2009, whose study of the radial variation of the rMS/bMS ratio
includes the results whose details we present here.)  Also, in the
following Section we will use the distance between the bMS and the rMS
to derive a definitive value for the helium abundance of the bMS.  (2)
The bMS appears to cross over the rMS and emerge on the other side of
it, in the subgiant region; we will discuss this further in Section
\ref{bothCMD}.  (3) A part of the white-dwarf sequence can be seen at
the lower left (but we will not discuss it in this paper).  (4) A
considerable number of undoubted cluster members lie too far above and
to the redward of the MS to be interpreted as binaries.  Although they
suggest some sort of sequence, their region is too ill defined to be
called a sequence.  This group too we will discuss in
Sect.\ \ref{bothCMD}, in conjunction with the CMD of the much richer
central region of the cluster.


%
\section{The helium abundance of the bMS}
\label{theo}
%

Ever since the surprising discovery that the bMS has higher metallicity
than the rMS (Piotto et al.\ 2005, P05), it has become increasingly
evident that this reversal of the usual color progression with MS
metallicity must indicate that the stars of the bMS contain a higher
proportion of helium.  Here we apply theoretical models of stellar
structure to the question of what helium abundance the location of the
observed blue sequence actually implies.

\subsection{Fitting the color separation of the bMS and the rMS}

For the comparison of observation with theory we select a single salient
characteristic of the bifurcation of the main sequence:\ the color
separation between the two branches.  We measure this at a magnitude
where the separation is large and the photometry is also quite reliable.
Examining Fig.\ 3, we see that within the magnitude range in which our
photometry is free from saturation the separation of the two sequences
appears to remain nearly constant over a stretch of about two
magnitudes, giving us a large enough number of stars to get a good value
of the color separation of the two sequences. We will refer to this
separation as $\Delta C$.  Specifically, we chose the magnitude interval
$20.6\le m_{\rm F606W}\le 21.8$, so that we can consider our observed
$\Delta C$ to apply to the entire middle part of this range, i.e., for
several tenths of a magnitude on either side of the midpoint, $m_{\rm
  F606W}=21.2$.

For the actual determination of the color separation we made use of a
procedure that is described in great detail by Bellini et al.\ (2009).
Briefly, we drew a fiducial color sequence along the gap between the two
branches, and made the sequences approximately vertical by subtracting
from the color of each star the fiducial color at its magnitude.  We
then plotted a histogram of the resulting colors, and fitted it with a
pair of Gaussians.  The separation that we find is $0.057\pm 0.0017$
mag, where the uncertainty is due only to the Poisson statistics of the
star numbers in the two sequences.  As a more conservative figure,
however, we prefer to assign a three-sigma uncertainty, and to quote a
color separation of $0.057\pm 0.005$.

For calculation of theoretical values of $\Delta C$ we chose absolute
magnitude 7.0.  For our distance modulus we took $(m-M)_0=13.70$ (Del
Principe et al.\ 2006), $E(B-V)=0.12$ (Harris 2010), and $A_V=3.1\,
E(B-V)$ (Cardelli et al.\ 1989); these give $(m-M)_V=14.072$, which we
rounded to 14.1, so that $M_{\rm F606W}=7.0$ corresponds to $m_{\rm
  F606W}=21.1$,
close to the middle of the magnitude range that we used for our
observational $\Delta C$.  (We note that the results that we derive
below are insensitive to our exact choice of distance modulus, since
the bMS and rMS run so closely parallel at these magnitudes.)

We wished to compare theoretical isochrones with the whole stretch of
main sequence that is shown in Figure 3,
and we therefore extended our models to lower masses ($M<0.5\,
M_\odot$).  We again took [$\alpha$/Fe] = +0.4, and used the physical
scenario described by Pietrinferni et al.\ (2004, 2006).  For these low
masses, we rely on the equation of state by Saumon, Chabrier, \& van
Horn (1995) for dense, cool matter, and on low-temperature opacities by
Ferguson et al.\ (2005) and high-temperature opacities by Rogers \&
Iglesias (1992).  The outer boundary conditions were fixed by adopting
the Next Generation model atmospheres provided by Allard et al.\ (1997)
and Hauschildt et al.\ (1999a,b).  We fixed the base of the atmosphere
at Rosseland optical depth $\tau=100$, i.e., deep enough for the
diffusion approximation to be valid.  We note also that although
detailed atmospheres are available only for canonical He abundances,
helium does not appreciably affect the atmosphere, because its only
effect would be on pressure-induced H$_2$--He absorption, which matters
only in stars of lower mass than we discuss here (F.\ Allard, private
communication).  At any given metallicity, the match between the more
massive models and the low-mass ones was made at a mass level where the
transition in luminosity and effective temperature between the two
regimes is smooth (usually $\sim0.5\, M_\odot$).

We computed stellar models for [Fe/H] = $-$1.62 and $-$1.32, to
represent the rMS and bMS respectively; and to allow for the
uncertainty of 0.2 dex in the relative metallicities of the bMS and
the rMS we also computed models with [Fe/H] = $-$1.52 and $-$1.12, to
represent alternative [Fe/H] values for the bMS.  For each value of
[Fe/H] for the bMS, we calculated models for
a set of helium abundances that reached beyond $Y=0.4$.
Within the heavy elements we used the $\alpha$-enhanced mixture of
Pietrinferni et al.\ (2006).  

For the transformation of the theoretical stellar models into the
observational plane we used the semi-empirical colors and bolometric
corrections of Pietrinferni et al.\ (2004), transformed according to
Appendix D of Sirianni et al.\ (2005).  We also verified, however, that
using the color--$T_{\rm eff}$ relations and bolometric corrections of
Hauschildt et al.\ (1999a,b) would produce practically the same result.
(In any case, whatever inaccuracies there may be in our transformation
are greatly reduced by the fact that the quantity that we use is the
{\it difference} between two colors.)


\begin{figure}[!ht]
\plotone{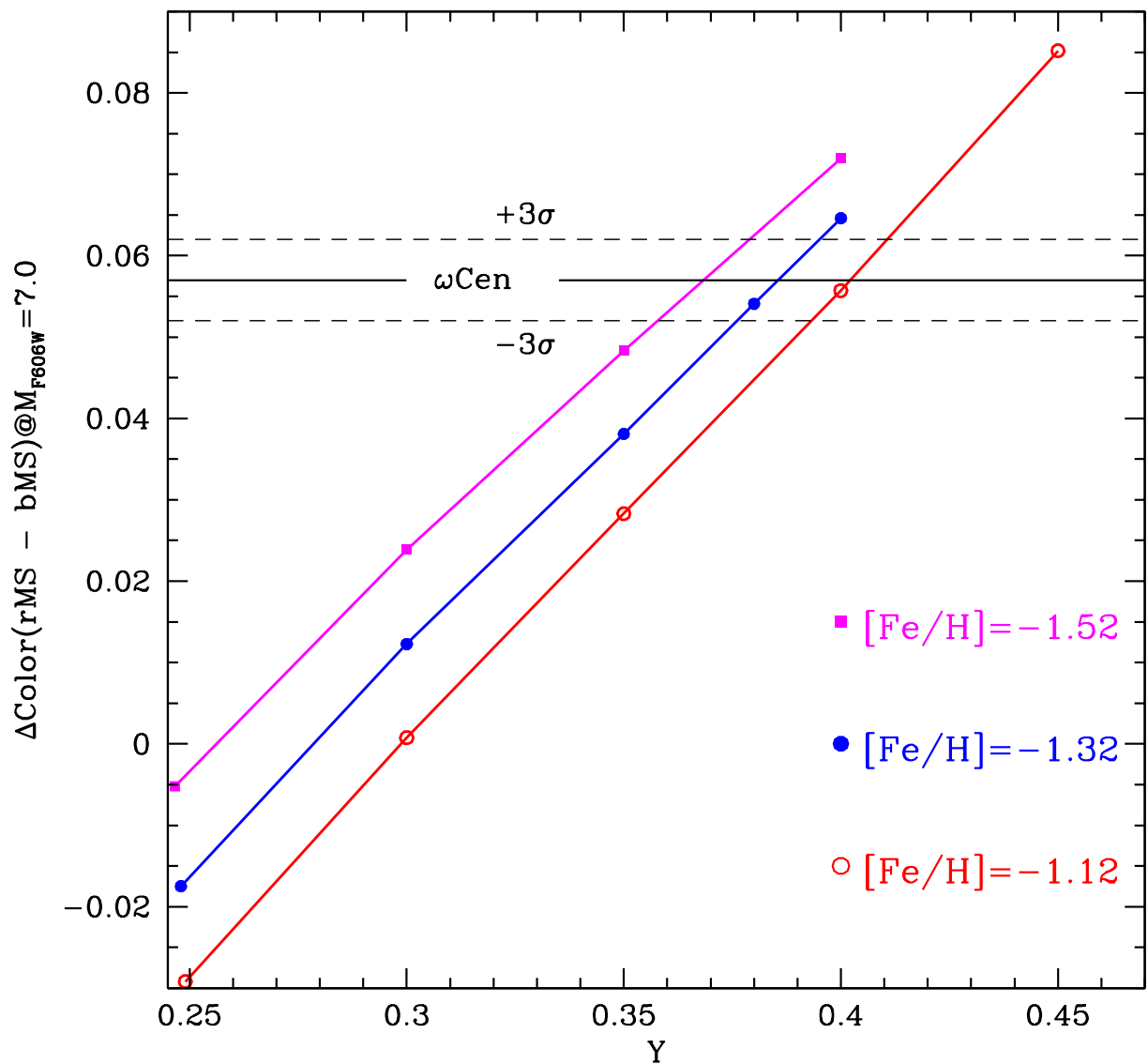}
\caption{Theoretical estimate of the parameter $\Delta C$ as a function
  of the initial He content of the bMS and for various assumptions about
  its iron content [Fe/H].  The horizontal
  solid line marks our measured value of $\Delta C$.  The horizontal
  dashed lines represent the 3$\sigma$ uncertainty of that
  measurement.}
\label{teoelio} 
\end{figure}

We used our theoretical models for a comparison with the observed
$\Delta C$ between the two branches of the main sequence.  To do this we
derived the equivalent theoretical quantity from our set of isochrones,
and transformed it into the observational plane, as just indicated.

The next step was to choose metallicity values for the rMS and for the
bMS, within the broad range of metal abundances that has been found in
spectroscopic studies --- a choice that is hampered, however, by the
lack of knowledge of how the two MS branches connect to the upper parts
of the HR diagram, from which all of our good abundance information
comes.  Thus all that we really have to go on is the uncertain result of
P05 that the metallicities of the rMS and the bMS are about $-$1.6 and
$-$1.3 respectively, with a somewhat less uncertain result that the
difference of the two metallicities is $0.3 \pm 0.2$.  (They italicize
the difference, marking it as their prime result concerning
metallicity.)

We accordingly chose for the rMS our [Fe/H] = $-$1.62 isochrone, with
primeval He (i.e., $Y=0.246$), and for the bMS the isochrones with
[Fe/H] = $-$1.32 and various He abundances.  In doing so, we note that
it is the {\it difference} of the metallicities that matters, rather
than the absolute value of either of them.  We could shift these
[Fe/H] values by 0.1--0.2 dex or so, without any appreciable effect on
the helium abundance that we will derive for the bMS, or its
uncertainty.  (We have, in fact, verified that our results remain the
same if we choose [Fe/H] = $-$1.75 for the rMS and also shift all the
bMS [Fe/H] values by the same 0.13.)

For comparison with observation, then, we calculated the color
difference at $M_{\rm F606W}=7.0$ between the rMS isochrone and the bMS
isochrones, with various assumed values of the helium abundance $Y$.
Bearing in mind, however, that P05 considered the metallicity difference
between the bMS and rMS to be uncertain by 0.2, we also carried out the
same procedure for assumed bMS metallicities of $-$1.12 or $-$1.52
(i.e., higher or lower by 0.2).

Figure 4 shows these theoretical $\Delta C$ estimates.  In order to
make our procedures clear, we explain at length the meanings of the
various lines in the figure: Each of the three sloping lines
corresponds to the bMS metallicity that is indicated for that line in
the color key.  Each point on the line shows the value of $\Delta C$
that comes from assuming that value of metallicity for the bMS, along
with a particular value of $Y$, while always using for the rMS [Fe/H]
= $-$1.62 and $Y=0.246$.  The lines themselves were created by
connecting with straight lines the points belonging to the same bMS
metallicity.

We used the figure to determine the value of the helium abundance $Y$,
and also its uncertainty.  The latter arises from two sources:\ the
uncertainty in the metallicity difference between the bMS and the rMS,
and our observational error in measuring $\Delta C$.

The solid horizontal line in the figure corresponds to our measured
value of $\Delta C$.  It serves two purposes: First, its intersection
with the sloping line for [Fe/H] = $-$1.32 (our preferred value for the
bMS) is at $Y$ = 0.0386; this is our result for the helium abundance.
Second, this line also tells us the uncertainty in $Y$ that results from
the uncertainty of 0.2 in the metallicity of the bMS relative to the
rMS.  To evaluate that uncertainty we simply measure the distance from
the intersection of the solid horizontal line with the [Fe/H] = $-$1.32
line to its intersection point with either the [Fe/H] = $-$1.12 or the
$-$1.52 line.  These distances are each 0.016, which is thus the
uncertainty in $Y$ due to this cause.

The second source of uncertainty in $Y$ comes from the measuring error
in $\Delta C$.  To represent this, we drew the two horizontal dashed
lines, which are above and below the solid line by the
$3\sigma=\pm0.005$ that we chose as a conservative error estimate.  They
intersect the sloping [Fe/H] = $-$1.32 line at $Y$ = 0.375 and 0.397
respectively, 0.011 greater or less than our $Y$ value of 0.386.

When we combine this $\pm$0.011 with the $\pm$0.016 that came from the
uncertainty in the metallicity difference, our result for the helium
abundance of the bMS is $Y = 0.39 \pm 0.02$.  Not only is this a more
reliable estimate than has been available before; it is the first
estimate that has had a quantitative uncertainty attached.

Recently, Pasquini et al.\ (2011) have used the He $\lambda$10830 line
in the spectra of two red-giant stars in NGC 2808 to estimate that the
He abundance of those stars is $Y\geq0.39$.  It is interesting to see
the similarity with what we find in $\omega$ Cen.  Additionally, in
$\omega$ Cen itself Dupree, Strader, \& Smith (2011) have detected the
$\lambda$10830 line in some RGB stars but not in others, in agreement
with our suggestion that some of the cluster stars have enhanced helium.

\subsection{The isochrones}

\begin{figure}[!ht]
\epsscale{1.0} \plotone{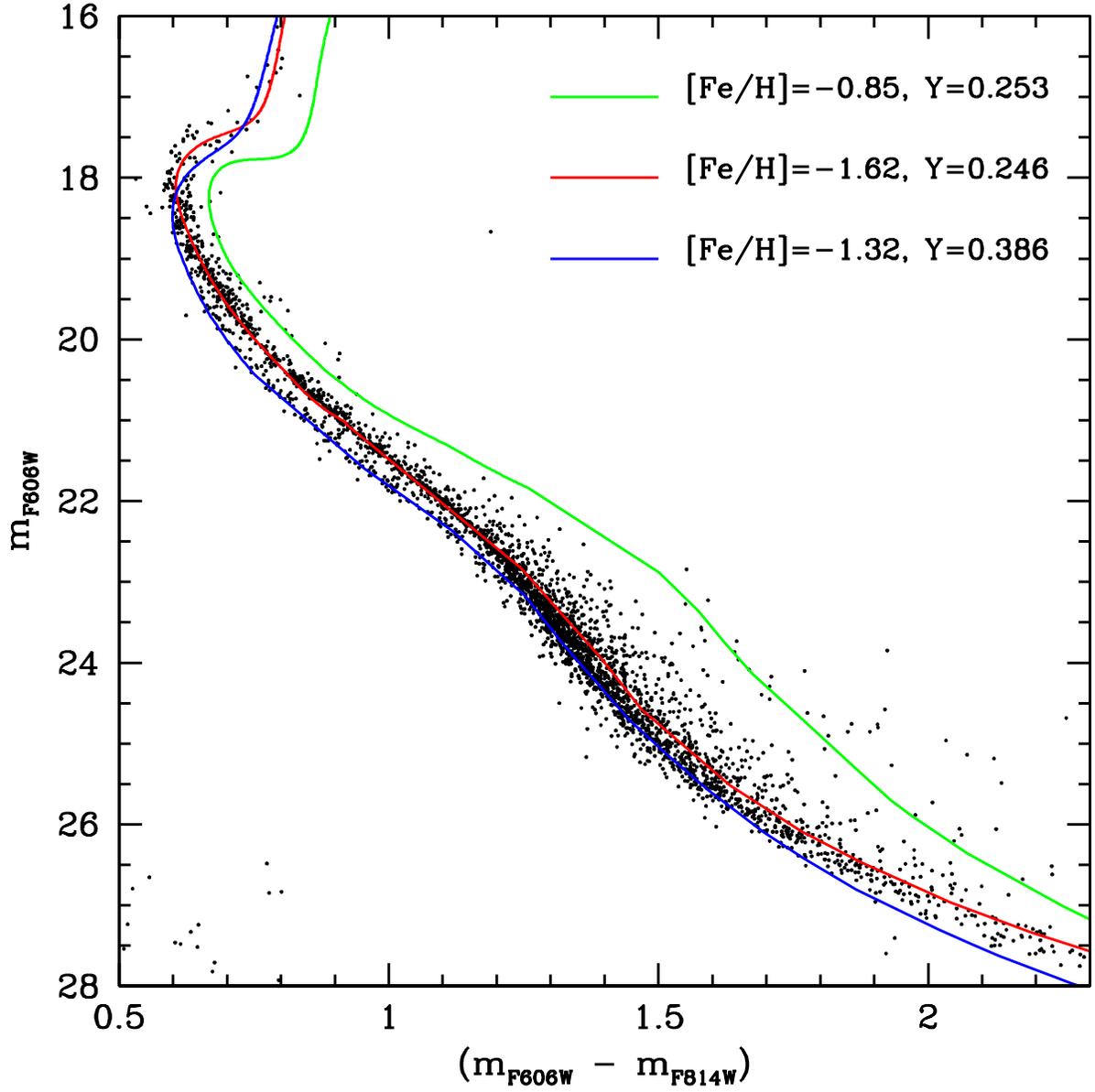}
\caption{Comparison of selected theoretical isochrones with the CMD of
  the outer field.  The isochrones drawn in red and in blue were
  chosen by merely matching the colors of the bMS and the rMS at
  $m_{\rm F606W}=21.17$.  The isochrone in green will be discussed in
  Section \ref{bothCMD}. }
\label{figteoobs} 
\end{figure}

Figure \ref{figteoobs} shows the isochrones that correspond to the
abundances with which we have fitted the color separation between the
bMS and the rMS (along with a third isochrone that will be discussed in
Section \ref{bothCMD}).  An isochrone with primeval helium and a
metallicity appropriate for $\omega$ Cen fits the rMS, while, as just
explained, we fit the bMS with an [Fe/H] that is higher by $\sim$0.3,
and helium content $Y=0.386$.  We find it gratifying that the
theoretical isochrones fit the bMS and the rMS as well as they do,
especially since the fitting is a direct overplot, with no adjustments
other than our having chosen the abundances so as to match the observed
colors at $m_{\rm F606W}=21.1$.

%
\section{A New Color-Magnitude Diagram for the Central Field}
\label{ctrCMD}
%

Our CMD of the outer field is photometrically accurate, and free of
field stars, but it has too few stars for us to draw clear conclusions
from it alone.  To illuminate what the outer field is telling us, we
will compare its CMD with that of the much richer central field, as
measured through the same pair of filters.  For this purpose, we improve
on the results of the {\it HST} Treasury Survey of globular clusters
(Sarajedini et al.\ 2007, Anderson et al.\ 2008), in two ways:\ first we
follow the suggestions of Anderson et al.\ (2008, Sect.\ 7.1) regarding
the use of quality indices to select a sample of the best-measured
stars, and then we correct the Treasury photometry of the selected stars
for differential reddening.  We do not make a proper-motion selection,
however, because the central field is so rich that only a tiny fraction
of the total number of stars are field stars, and restricting our
attention to a proper-motion sample would have caused us to lose all the
faint stars.

   \begin{figure}[!ht]
   \centering
   \plotone{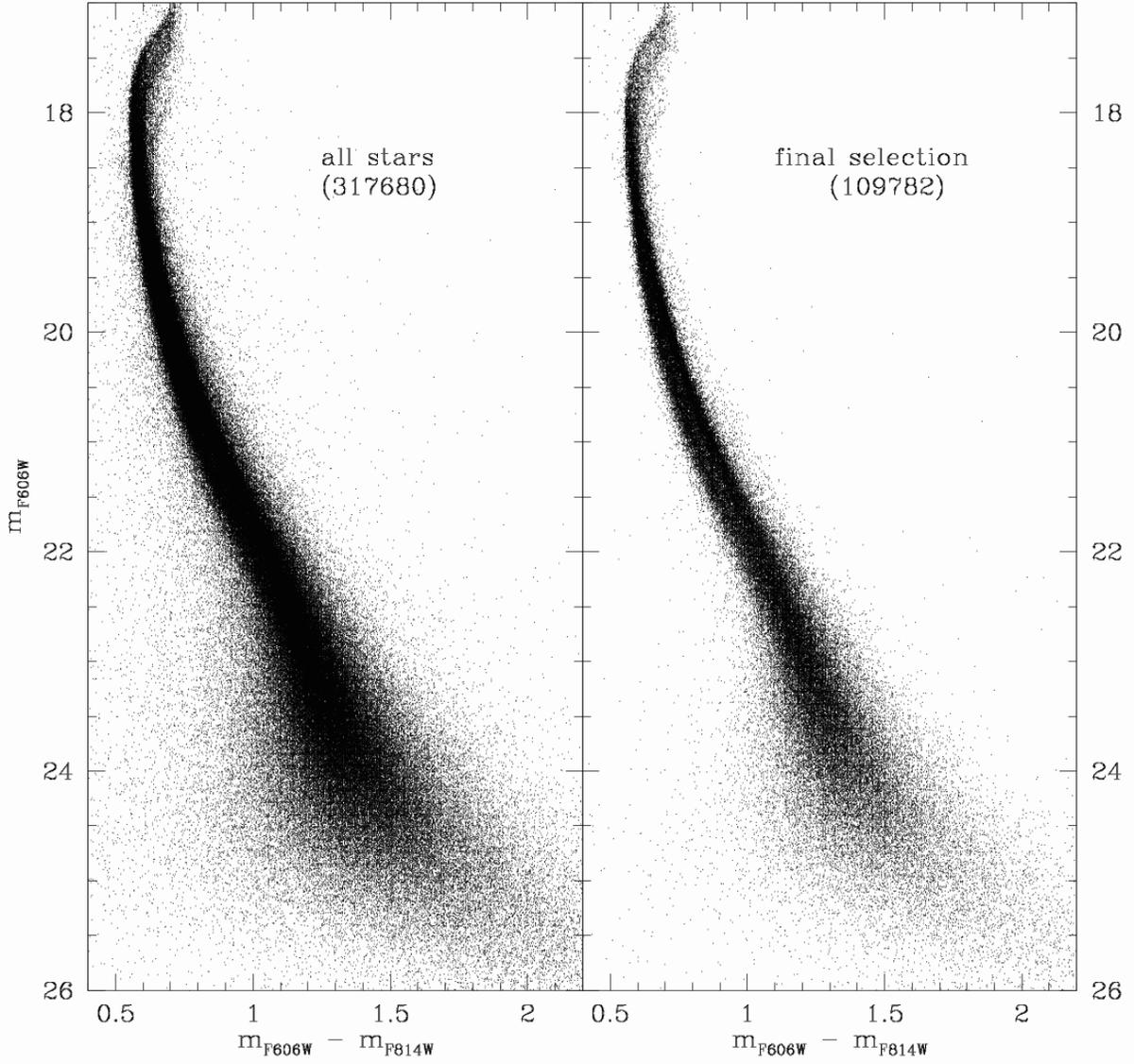}
   \caption{Comparison of a portion of the CMD of the central field,
     before and after selection of the stars that are the most likely to
     be photometrically reliable.}
         \label{befaft}
   \end{figure}

\subsection{Photometric Quality Selection}
\label{qualsel}

Our photometric quality criteria were four in number: The first two were
the rms residuals of the magnitude measures in each band (columns headed
``err'' in the Globular Cluster Treasury files at 
http://www.astro.ufl.edu/$\sim$ata/public\_hstgc/), while the other two
were {\tt xsig} and {\tt ysig}, which are not internal errors but rather
the differences between the mean positions measured in F606W and in
F814W, for the $x$ and $y$ coordinates respectively.  (Experience in
working with the Treasury results had shown that stars with low-quality
photometry tended also to have larger values of {\tt xsig} and {\tt
ysig}.)  We plotted each criterion against magnitude, and then tried
rejecting stars at various percentile levels of the criteria.  We found
that a good compromise between improving the CMD, on the one hand, and
rejecting too many stars on the other hand, was to reject stars that
were above the 75th percentile of any of the four criteria.  This
process, illustrated in Figure \ref{befaft}, selected 109,782 of the
317,680 stars that had at least two deep exposures in each filter.  The
cut is severe, but that is understandable in a field as crowded as the
center of $\omega$ Centauri, and we believe that the resulting CMD is as
good as we can produce for this field by quality selection alone.

(As a test of the efficacy of the individual criteria we verified that
each of them rejected numerous stars that had passed all the other
criteria at the 75th percentile level.  We note that we also tried
four additional criteria, which measure the quality of the PSF fit and
the amount of encroaching light from neighboring stars, for each
filter --- called {\tt qv}, {\tt qi}, {\tt ov}, and {\tt oi} by
Anderson et al.\ 2008.  We found that these criteria rejected few
stars that were not already flagged by the four that we did use; we
concluded that such low levels of rejection could be attributed to
random scatter of the values of those criteria, and we therefore did
not use them at all.)

\begin{figure}[!hb]
\epsscale{1.00}
\plotone{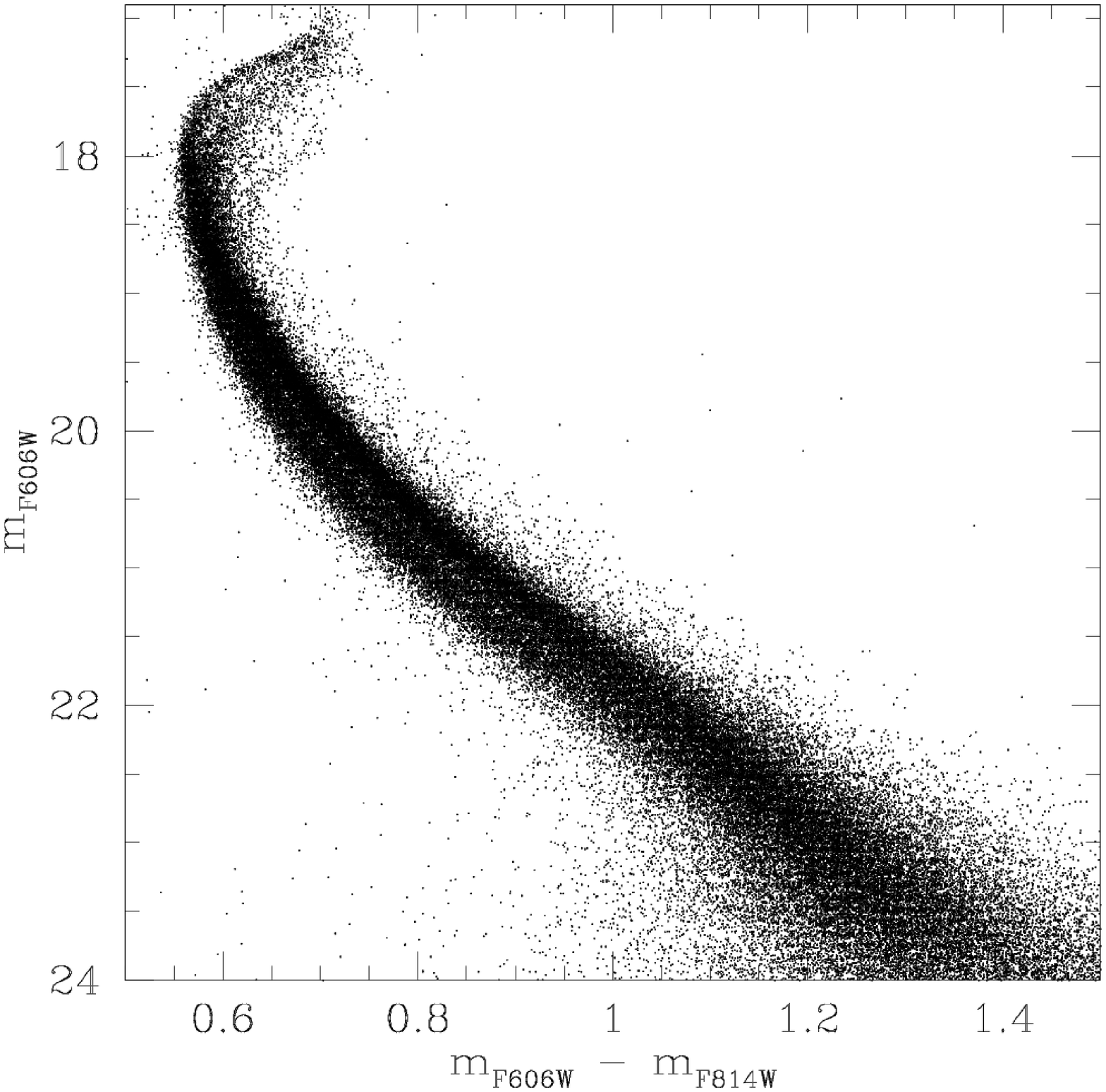}
\caption{The CMD of the central field, after quality selection and
    reddening correction.}
\label{cCMDfig}
\end{figure}

\begin{figure}[!hb]
\epsscale{1.00}
\plotone{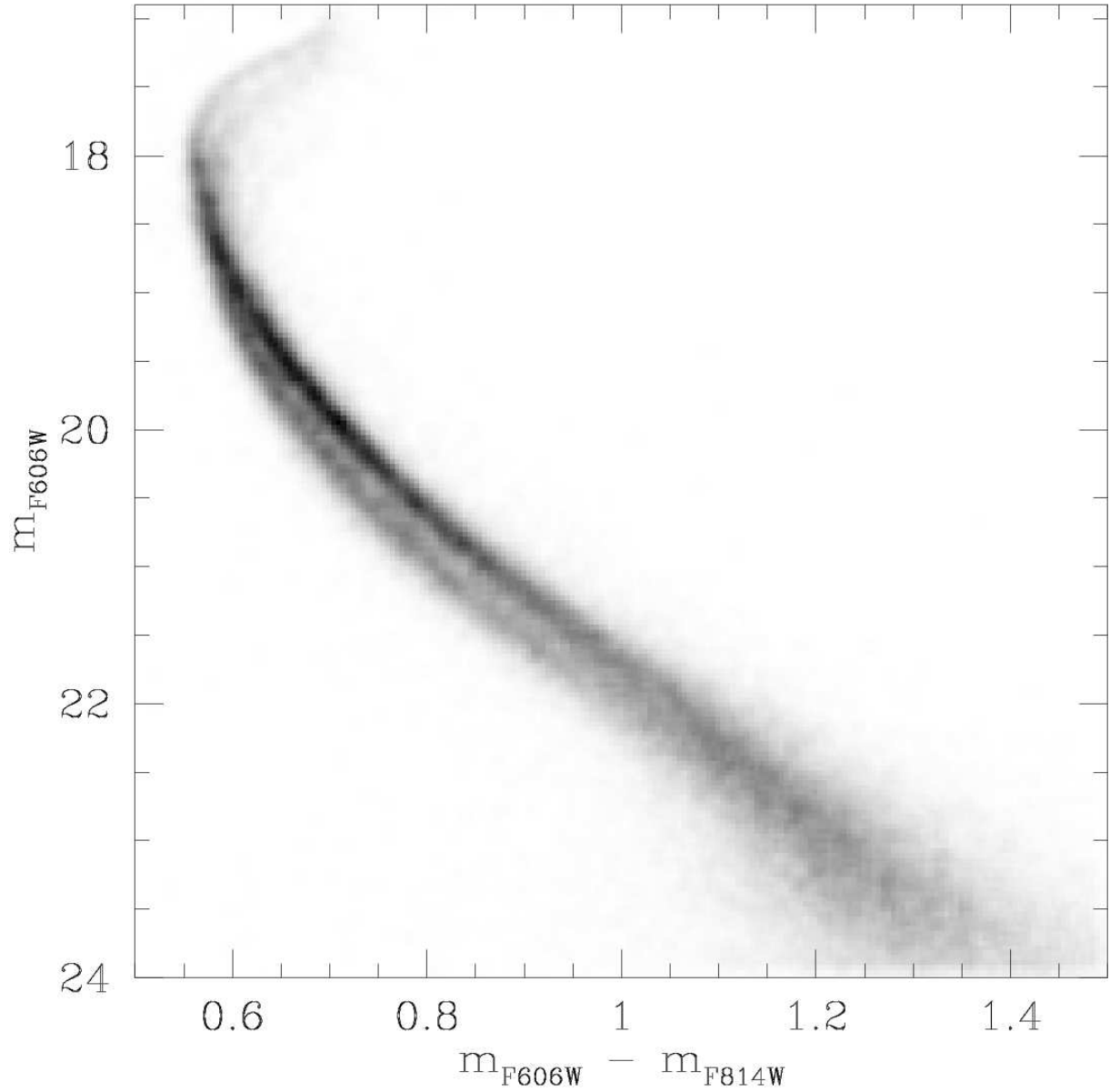}
\caption{The CMD of the central field, after quality selection and
  reddening correction, shown as a Hess diagram.  This gray-scale
  representation was made from the same colors and magnitudes that were
  shown in Fig.\ \ref{cCMDfig}.}
\label{hess}
\end{figure}

\subsection{Corrections for Differential Reddening}
\label{redcor}

The sequences in the CMD of $\omega$ Cen are somewhat broadened by
differential reddening.  The Harris (2010) catalog gives for this
cluster $E(B-V)=0.12$, and reddening as large 
as this is rarely uniform.  The basic method that we used to construct a
reddening map of our field depends on drawing a fiducial sequence to
follow the course of the MS, SGB, and RGB, and then deriving an
approximate reddening from the observed color of each star, by seeing
how far it needs to be slid up or down a reddening line in order to meet
the fiducial sequence.  Our usual procedure would then be to designate
the well-observed stars in a bright interval of magnitude as reference
stars, and to take for the reddening correction of each individual star
the mean reddening of the 75 nearest reference stars.  For $\omega$ Cen,
however, there are special problems, because the multiplicity of
sequences confuses the procedure.  In this case we proceeded as follows:

We began, as we have for other clusters, by rotating the CMD through an
angle
$$\theta = \tan^{-1}
            {A_{\rm F606W}\over A_{\rm F606W}-A_{\rm F814W}} \, , $$
around an arbitrarily chosen point.  This rotation makes the reddening
line into the new $x$ axis, greatly simplifying the de-reddening
operation, which is now just a shift along the $x$ axis.

For the special problem of $\omega$ Cen, we iterated the first part of
our usual procedure.  As a first approximation to a fiducial sequence,
we arbitrarily drew a line along the rMS, simply because it is the more
populous of the MS components.  We then applied our usual procedure,
even though it is imperfect because of the multiple sequences; it did
produce enough differential-reddening correction to effect some
narrowing of the MS.  (Because the rMS is so dominant numerically, any
attempt to move bMS stars onto the rMS merely introduces noise that
slows the convergence.)  Four iterations of the correction procedure
sufficed to give as good a correction as we could get.

In the corrected CMD (Fig.\ \ref{cCMDfig}) the subgiant region is split
into multiple branches (a foretaste of what we may hope to find with a
richer set of filters).  The MS split has become less prominent here,
but only as a result of printing the CMD heavily enough to show the
details of the subgiant region.  Fig.\ \ref{hess} is a Hess diagram made
from the same colors and magnitudes.  The split in the main sequence can
be seen more clearly now --- strikingly well, in fact, for the crowded
center of the cluster.

%
\section{Superposition of the Two Color-Magnitude Diagrams}
\label{bothCMD}
%

We now have a CMD of the outer field, quite sharp and almost completely
free of field stars, but with few stars overall; and we have a CMD of
the central field in the same two filters, with a much larger number of
stars but with photometry of lower quality, in that crowded region.  A
fruitful step now is to compare the two CMDs by laying one on top of the
other.  In order to make this superposition, however, we need to adjust
the zero points of the reddening-corrected magnitudes, because the
corrections did not preserve zero points.  We did this adjustment by
eye, simply by getting the best match that we could.

\begin{figure}[!ht]
\epsscale{0.80}
\plotone{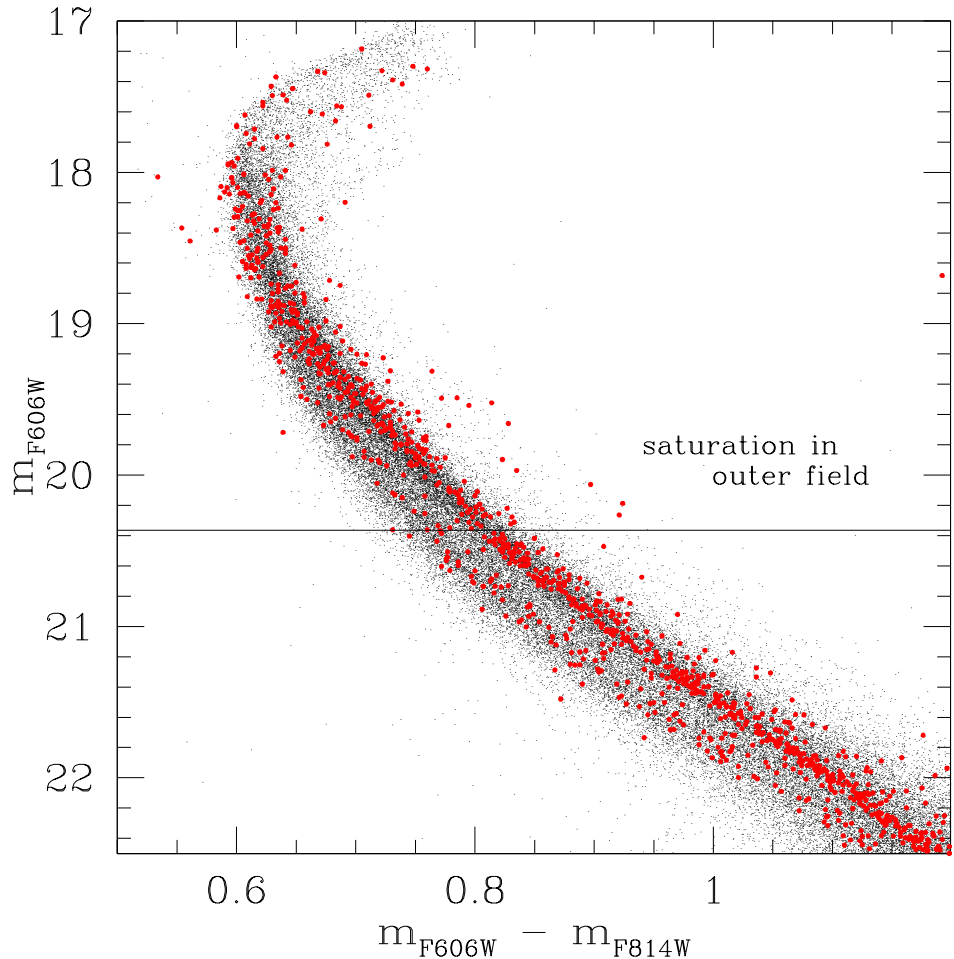}
\caption{The points from the outer field, in red, superposed on points
  from the central field, in black.  This figure shows the bright part
  of the magnitude range covered by our observations, while
  Fig. \ref{fitoutft} shows the faint part of the range (with
  considerable magnitude overlap between the two).}
\label{fitoutbr}
\end{figure}

\begin{figure}[!ht]
\epsscale{0.80}
\plotone{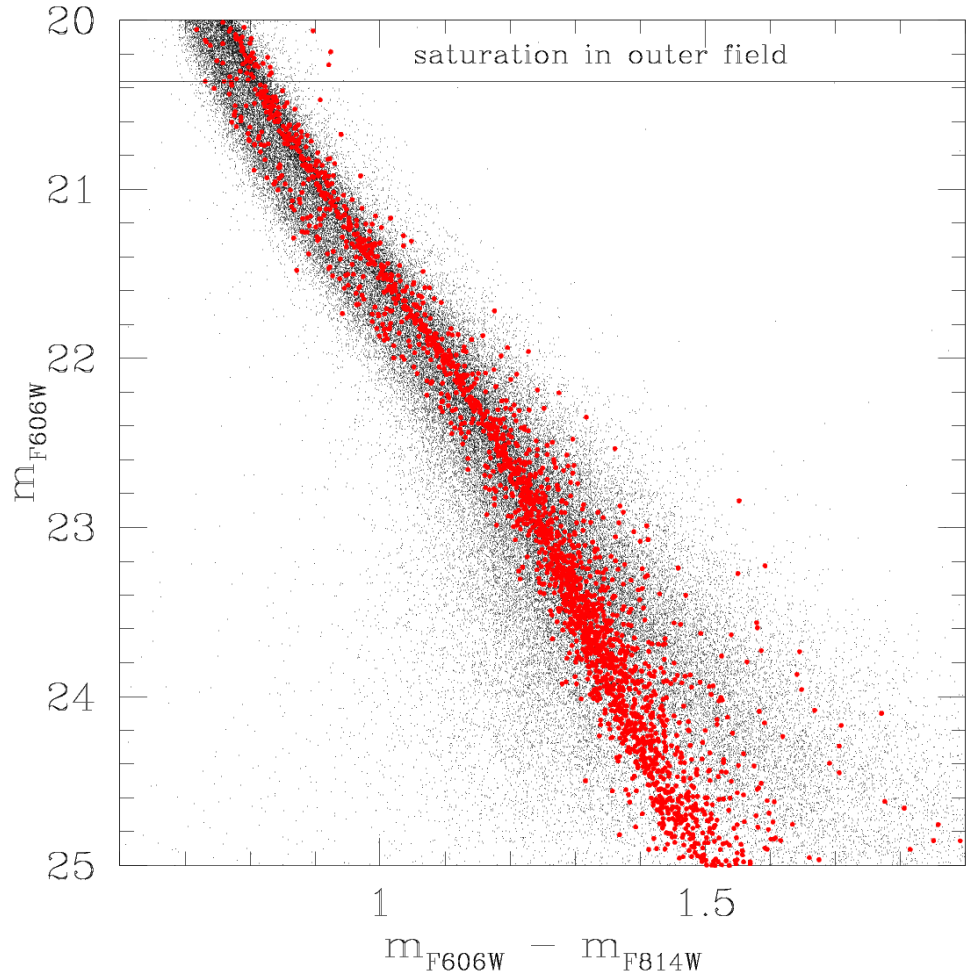}
\caption{The faint end of the superposition of the CMD of the outer
  field (red) on that of the central field (black).}
\label{fitoutft}
\end{figure}

The superposition, which now allows us to combine the clarity and purity
of the outer field with the richness and continuity of the sequences
that we see in the central field, is shown in Figures \ref{fitoutbr} and
\ref{fitoutft}.  The first of them shows the bright part of the
magnitude range, and the second shows the faint part (with considerable
magnitude overlap between the two).

In Fig.\ \ref{fitoutbr}, the rMS appears to continue upward into the
bright edge of the subgiant-branch (SGB) stars, while the bMS seems to
emerge from the crossing of the bMS and rMS at a magnitude level that
takes it into the next-most-prominent of the SGB sequences.
Although a result that depends on photometry of saturated star images
must always be taken with some reserve, it is hard to see how any other
connection of branches of the MS with branches in the SGB region could
be possible, given the numbers of stars along each of the two sequences,
on the MS and then in the SGB region.  (It is interesting to note
that the detailed study of Villanova et al.\ 2007, which was limited
to ground-based resolution beyond the central $10^\prime \times
10^\prime$ of the cluster, made SGB connections for the rMS and for the
reddest branch of the MS [which they called MS-a], but they were unable
to make a connection for the bMS.  It is our superposition of the CMD of
the outer field on that of the central field that has allowed us to
suggest a SGB connection for the bMS.)

In Figure \ref{fitoutft} we show the fainter part of the superposed
CMDs.  Here the bMS seems to intersect the rMS at magnitude $\sim
22.8$, but it is not at all clear whether this is a crossing or a
merger.  There is a faint hint that the bMS might emerge on the red
side of the rMS at fainter magnitudes, but it is impossible to be sure
of this, because the photometry in the central field rapidly loses
accuracy, and below $m_{\rm F606W}\sim 23$ photometric error begins to
spread the MS hopelessly.

We have already called attention to the group of faint red stars well to
the red of the MS, in the outer field; we see now that they are
prominent in the central field too.  Although we have given up hope of
tracing sequences at such faint magnitudes, what we can say nevertheless
is that the magnitude spread is clearly greater on the red side of the
MS than on the blue side, so that this redward extension must be
something real.

We have noted that the faint red stars cannot be explained away as
binaries, since they were shown in Fig.~\ref{outfig} to be well above
the upper limit of the region in the CMD where binaries lie.  Since a
redder MS color is usually associated with a higher metallicity (with
the notable exception of the bMS!), these stars might be suspected to be
the lower end of the faintest SGB sequence (the region that Villanova et
al.\ 2007 called MS-a).  This component of $\omega$ Cen is known to
be much less metal-poor, with [Fe/H] $-$0.8 to $-$0.7 (Johnson \&
Pilachowski 2010).  In Fig.~\ref{figteoobs} we have chosen from our
library of isochrones the one with [Fe/H] = $-$0.85, $Y=0.253$, and age
13 Gyr.  (Noting that the highly enhanced helium in the bMS and in a few
stars in NGC 2808 is an unusual anomaly, we chose to revert to a more
normal $Y$, raising its value just a little as a consequence of the
increased metallicity, and using a conventional value of $dY/dZ$.)

On the other hand, the spread of the red stars in color is far greater
than can be expected in any single sequence, nor can we confidently
trace in the observed CMD a sequence that connects SGB-a with these
stars --- and in fact in Villanova et al.\ (2007) we suggested
tentatively that MS-a continues down into the rMS at a brighter
magnitude than this.  It is unfortunate that the outer field does not
have enough stars to allow the tracing of any but the two richest
sequences, while at faint magnitudes the photometry of the central field
is too damaged by crowding to yield further enlightenment.

With observations that are confined to these two filters, we can go no
farther in tracing sequences.  It has recently been shown by Bellini et
al.\ (2010), however, that when observations with the ultraviolet
filters of WFC3 are brought into the picture, all sorts of details
emerge.  We will therefore leave further pursuit of the continuity of
sequences to future papers that include results from more filters.

%
\section{Summary}
\label{summ}
%

We have used \hst\ ACS/WFC imaging of an outer field of $\omega$
Centauri at two epochs to derive a color-magnitude diagram, cleaned of
field stars, that shows the clearest separation of the blue and red
branches of the main sequence that has yet been achieved.  
We have calculated new stellar-structure models
for a mesh of abundances of helium and metals, and have fitted
isochrones to our observed sequences, deducing for the blue branch of
the main sequence $Y=0.39\pm 0.02$ --- the first time that this
anomalous abundance of helium has been given a solid base in theory, and
assigned a quantitative uncertainty.  We show a plausible fit of the
$\omega$ Cen sequences with three different isochrones that have
different abundances of helium and metals.

We have re-examined the CMD of the central part of $\omega$ Cen,
selected a subset of stars that are likely to have the most reliable
photometry, and corrected its photometry for differential reddening.
When we superpose the CMDs of the two regions, we tentatively identify
the SGB continuation of the bMS with one of the many branches that the
rich central field shows in the SGB region.  We also find a striking new
group of stars to the red of the main sequence.  We look forward to new
insights from combining these results with the leverage that UV
photometry with WFC3 now applies to this extraordinarily complex CMD.

\acknowledgements
S.C.\ warmly thanks F.\ Allard for interesting discussions and for
kindly sharing her own results. S.C., A.P., and G.P.\ acknowledge
partial support by MIUR under grants PRIN2007 (prot.\ 20075TP5K9) and
PRIN-INAF 2009, and G.P.\ acknowledges support by ASI under grants
ASI-INAF I/016/07/0 and I/009/10/0.
J.A.\ and I.R.K.\ acknowledge support from STScI grants GO-9444,
GO-10101, and GO-11233.\\ 
%

\begin{center}
\textbf{REFERENCES}
\end{center}
\begin{description} 
\item Allard F., Hauschildt, P.~H., Alexander, D.~R., \& Starrfield, S.M.,  1997, \araa, 35, 137
\item Anderson, J. 1997, Ph.D. thesis, Univ. of Calif., Berkeley
\item Anderson, J. 2002, in Proceedings of the 2002 HST Calibration Workshop, ed.\ S. Arribas, A. Koekemoer, \& B. Whitmore (Baltimore:\ STScI), p.\ 13
\item Anderson, J. 2006, in Proceedings of the 2005 Calibration Workshop, ed.\ A.\ Koekemoer, P.\ Goudfrooij, \& L.\ Dressel (Baltimore:\ STScI), p.\ 11
\item Anderson, J., \& King, I.\ R. 2006, ACS Instrument Science Report 2006-01 
\item Anderson, J., et al. 2008, \aj, 135, 2055 
\item Bedin, L. R., Piotto, G., Anderson, J., Cassisi, S., King, I.\ R., Momany, Y., \& Carraro, G. 2004, \apjl, 605, L125
\item Bedin, L.\ R., 2005, Cassisi, S., Castelli, F., Piotto, G., Anderson, J., Salaris, M., Momany, Y., \& Pietrinferni, A. \mnras, 357, 1038	
\item Bellini, A., Piotto, G., Bedin, L.\ R., King, I.\ R., Anderson, J., Milone, A.\ P., \& Momany, Y. 2009, A\&A, 507, 1393
\item Bellini, A., Bedin, L.\ R., Piotto, G., Milone, A.\ P., Marino, A.\ F., \& Villanova, S. 2010, \aj, 140, 631
\item Bragaglia, A., Carretta, E., Gratton, R., D'Orazi, V., Cassisi, S., \& Lucatello, S. 2010a, A\&A, 519, A60
\item Bragaglia, A., Carretta, E., Gratton, R.\ G., Lucatello, S., Milone, A., Piotto, G., D'Orazi, V., Cassisi, S., Sneden, C., \& Bedin, L. R. 2010b, ApJ, 720, L41
\item Cannon, R. D., \& Stobie, R. S. 1973, MNRAS, 162, 207 
\item Cardelli, J. A., Clayton, G. C., \& Mathis, J. S. 1989, ApJ, 345, 245 
\item D'Antona, F., \& Caloi, V. 2008, MNRAS, 390, 693
\item Del Principe, M., et al. 2006, \apj, 652, 362 
\item Dickens, R. J., \& Woolley, R. v. d. R. 1967, R. Obs. Bull., 128, 255  
\item Dupree, A.\ K., Strader, J., \& Smith, G.\ H. 2011, ApJ, 728, 155 
\item Ferguson, J.\ W., Alexander, D.\ R., Allard, F., Barman, T., Bodnarik, J.\ G., Hauschildt, P.\ H., Heffner-Wong, A., \& Tamanai, A. 2005, \apj, 623, 585
\item Freeman, K. C., \& Rodgers, A. W. 1975, \apj, 201, L71
\item Gilliland, R. 2004, ACS Instrument Science Report 2004-01
\item Gratton, R.\ G., Carretta, E., Bragaglia, A., Lucatello, S., \& D'Orazi, V. 2010, A\&A, 517, A81
\item Harris, W.\ E. 2010, arXiv 1012.3224
\item Hauschildt, P.\ H., Allard, F., \& Baron, E. 1999a, \apj, 512, 377
\item Hauschildt, P.\ H., Allard, F., Ferguson, J., Baron E., \& Alexander, D. 1999b, \apj, 525, 871
\item Johnson, C.\ I., \& Pilachowski, C.\ A.  2010, \apj. 722, 1373
\item Lee, Y.-W., Joo, J.-M., Sohn, Y.-J., Rey, S.-C., Lee, H.-C., \& Walker, A. R. 1999, Nature, 402, 55 
\item Merritt, D., Meylan, G., \& Mayor, M. 1997, \aj, 114, 1074
\item Norris, J. 2004, \apj, 612, L25
\item Norris, J. E., Freeman, K. C., \& Mighell, K. J. 1996, \apj, 462, 241
\item Pancino, E., Ferraro, F., Bellazzini, M., Piotto, G., \& Zoccali,  M.  2000, \apj, 534, 83 
\item Pasquini, L., Mauas, P., K\"aufl, H.\ U.,\& Cacciari, C. 2011, A\&A, 531, A35
\item Pietrinferni, A., Cassisi, S., Salaris, M., \& Castelli, F. 2004, \apj, 612, 168
\item Pietrinferni, A., Cassisi, S., Salaris, M., \& Castelli, F. 2006, \apj, 642, 797
\item Piotto, G., et al.\ 2005, \apj, 621, 777 
\item Piotto, G., et al.\ 2007, \apj, 661, L53 
\item Rogers, F.\ J., \& Iglesias, C.\ A. 1992, \apjs, 79, 507
\item Sarajedini, A., et al.\ 2007, \aj, 133, 1658 
\item Saumon, D., Chabrier, G., \& van Horn, H.\ M. 1995, \apjs, 99, 713
\item Sirianni, M., et al.\ 2005, \pasp, 117, 1049 
\item Sollima, A., Pancino, E., Ferraro, F.\ R., Bellazzini, M., Straniero, O., \& Pasquini, L. 2005, \apj, 634, 332
\item Villanova, S., et al.\ 2007, \apj, 663, 296 
\end{description}

%
\end{document}